\documentclass[
reprint,
superscriptaddress,
amsmath,amssymb,
aps,
prx,
nofootinbib
]{revtex4-2}

\usepackage{graphicx} 
\usepackage{hyperref}
\usepackage{lipsum}
\usepackage{siunitx}
\usepackage{physics}
\usepackage{color}
\usepackage[capitalise]{cleveref}
\usepackage{verbatim}
\usepackage[version=4]{mhchem}
\DeclareSIUnit[]{\cps}{\text{counts/s}}

\newcommand\blfootnote[1]{%
	\begingroup
	\renewcommand\thefootnote{}\footnote{#1}%
	\addtocounter{footnote}{-1}%
	\endgroup
}

\begin{document}

	\title{Heralded entanglement of on-demand spin-wave solid-state quantum memories for multiplexed quantum network links}
	\author{Jonathan Hänni}
	\thanks{These authors contributed equally.}
	\author{Alberto E. Rodríguez-Moldes}
	\thanks{These authors contributed equally.}
	\author{Félicien Appas$^\ddagger$}
	\thanks{These authors contributed equally.}
	\author{Soeren Wengerowsky}
	\author{Dario Lago-Rivera}
	\thanks{Present address: TOPTICA Photonics AG, Lochhamer Schlag 19, 82166
		Gräfelfing, Germany}
	\author{Markus Teller}
	\author{Samuele Grandi}
	\affiliation{ICFO - Institut de Ciencies Fotoniques, The Barcelona Institute of Science and Technology, Castelldefels (Barcelona) 08860, Spain}
	\author{Hugues de Riedmatten}
	\affiliation{ICFO - Institut de Ciencies Fotoniques, The Barcelona Institute of Science and Technology, Castelldefels (Barcelona) 08860, Spain}
	\affiliation{ICREA – Institucio Catalana de Recerca i Estudis Avançats, 08015 Barcelona, Spain}

	\date{\today}
	\begin{abstract}
	 The ability to distribute heralded entanglement between distant matter nodes is a primitive for the implementation of large-scale quantum networks.
    Some of the most crucial requirements for future applications include high heralding rates at telecom wavelengths, multiplexed operation and on-demand retrieval of stored excitations for synchronization of separate quantum links. Despite tremendous progress in various physical systems, the demonstration of telecom-heralded entanglement between quantum nodes featuring both multiplexed operation and on-demand retrieval remains elusive. In this work, we combine narrowband parametric photon-pair sources and solid-state quantum memories based on rare-earth doped crystals to demonstrate telecom heralded entanglement between spatially separated spin-wave quantum memories with fully adjustable recall time and temporal multiplexing of 15 modes. In a first experiment, the storage in the spin-state is conditioned on the entanglement heralding.  
    We take advantage of the control over readout pulse phase to achieve feed-forward conditional phase-shifts on the stored photons depending on which heralding detector clicked. We exploit this effect to double the entanglement heralding rate for a given quantum state up to \SI{510}{\cps}, with an associated detection rate of \SI{0.32}{\cps} and measured positive concurrence by up to 6 standard deviations. In a second experiment, we  simulate the communication time of a long-distance link by implementing an unconditional storage scheme with a dead-time of \SI{100}{\micro\second}. We take advantage of temporal multiplexing to increase the entanglement rates by a factor of 15 with respect to single mode storage, reaching a value of \SI{22}{\cps} per heralding detector. These results   establish our architecture as a prime candidate for the implementation of scalable high-rate quantum network links.
    \end{abstract}
    
	\maketitle
	
	\section{Introduction}
\blfootnote{$^\ddagger$ Corresponding author: felicien.appas@icfo.eu.}
The realization of a large-scale network of interconnected quantum nodes is one of the most promising avenues of quantum technologies \cite{Wehner2018, Kimble2008}. In this vision, matter qubits at remote locations are linked by single photons traveling in optical fibers, which distribute non-classical correlations that can be used as a resource for a number of compelling applications. Near-term quantum-assisted distributed tasks include blind quantum computing or quantum leader election~\cite{Wehner2018} ultimately leading to protocols taking advantage of the full potential of a quantum network such as distributed quantum computing \cite{Jiang2007}, distributed sensing~\cite{Zhang2021} or quantum cryptography~\cite{Pirandola2020}.
One of the challenges that such an endeavor has to face is the exponential absorption losses these photons have to withstand when traveling over long distances in optical fibers, an effect that cannot be counteracted by cloning or amplifying due to the intrinsic quantum nature of the light field.
Quantum repeaters have been proposed as a way to overcome this limitation 
by splitting long links into smaller segments, where entanglement between quantum memories (QM) placed at the edges can be established in a heralded way \cite{Briegel1998,Duan2001}. Once two adjacent segments are entangled, a swapping operation can be performed on the two neighboring QMs to extend the remote entanglement over the entire network.
The scalability of this architecture over large distance requires high entanglement generation rates and a quantum memory with high efficiency and long storage time. Other key enabling features are the ability to retrieve the stored quantum state on-demand and to store multiple photonic modes in a single QM. The former serves as an essential tool for synchronizing swapping operations between adjacent segments while the latter is mandatory to achieve high enough rates for a network application by allowing to perform continuous entanglement creation attempts during the latency introduced by the finite travel time of photons between nodes \cite{Simon2007}. 
Heralded entanglement of spatially separated atomic systems has already been performed in a variety of physical systems including single neutral atoms and ions \cite{Ritter2012,vanLeent2022,Krutyanskiy2023}, color centers in the solid-state \cite{Bernien2013, Knaut2024,Stolk2024} or cold atomic clouds~\cite{Chou2005, Liu2024} but their multiplexing capabilities have so far remained limited, hindering their use in high-rate long-distance quantum links.

In this work, we bridge the gap between previous realizations in different platforms by achieving the demonstration of telecom-heralded entanglement of ensemble-based quantum memories featuring both on-demand and multiplexed storage. Our implementation is based on spin-wave atomic frequency comb (AFC) storage in \ce{Pr^3+}:\ce{Y_2SiO_5} rare-earth ion doped crystal (REIC) QMs and non-degenerate cavity-enhanced spontaneous parametric downconversion (cSPDC) sources. 
This system benefits from multiplexed storage in several degrees of freedom such as temporal \cite{Afzelius2009}, spectral \cite{Sinclair2014,Seri2019} or spatial \cite{Gundogan2012, Teller2025} as well as the possibility to achieve high storage efficiency \cite{Hedges2010,Afzelius2010b,Duranti2024} and ultra-long storage times \cite{Zhong2015,Holzapfel2020,Ma2021}. Crucially, all of these features are compatible with on-demand storage and retrieval from the QM \cite{Seri2017}. Heralded memory-memory entanglement with predetermined storage time \cite{Usmani2012,Lago-Rivera2021, Liu2021} has already been established using REICs. However the high level of noise generated by the strong optical control pulses needed to perform on-demand storage has thus far hindered the demonstration of this functionality in entanglement distribution between quantum nodes.
Thanks to a significant improvement in signal-to-noise ratio for single photon storage with respect to the state of the art, we are able to demonstrate telecom-heralded entanglement between two spin-wave quantum memories with on-demand read-out by up to six standard deviation and for storage times of up to \SI{25}{\micro\second}. We exemplify the versatility of our storage protocol by performing, for the first time in rare-earth solid-state QMs, conditional state manipulation upon photon retrieval by active feedback on the readout pulse phase, leading to heralding rates up to 510 events/s and detection rates of \SI{0.32}{\cps}. Finally, we highlight the linear increase of the heralding rate with the number of stored temporal modes in an unconditional entanglement generation sequence, with recorded values as high as 22 events/s per idler detector.

\section{Experimental Setup}
\label{sec:setup}
\begin{figure*}
    \centering
    \includegraphics[width=\linewidth]{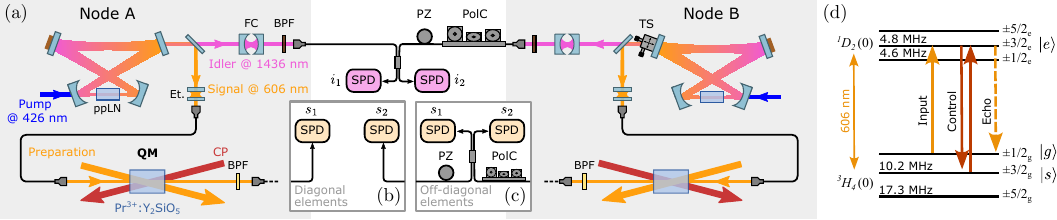}
    \caption{(a) Experimental setup. Each quantum node consists of a non-degenerate cSPDC source and a \ce{Pr^3+}:\ce{Y_2SiO_5} solid-state QM. Heralding is performed at an intermediate station by mixing the idler fields on a 50/50 beam splitter and recording single photon detection events, which herald an entangled state of a delocalized excitation shared across the two QMs. For density matrix reconstruction, single photons retrieved from the memory crystals are either (b) directly detected (diagonal elements measurement) or (c) interfered at a 50/50 BS (off-diagonal elements measurement). ppLN: Periodically Poled Lithium Niobate. FC: Filter Cavity. BPF: Bandpass Filter. PZ: Piezoelectric Fiber Stretcher. PolC: Polarization Controller. QM: Quantum Memory. CP: Control Pulse. SPD: Single Photon Detector. TS: Translational Stage. Et: Etalon Filter. (d) Energy levels corresponding to the the $^{3}H_4(0) \leftrightarrow ^{1}D_2(0)$ transition in \ce{Pr^3+}:\ce{Y_2SiO_5}. Input signal photons at \SI{606}{\nano\metre} from the cSPDC sources are resonant with the $1/2_g\leftrightarrow3/2_e$ transition, where they are absorbed. The optical excitation is then stored into and then retrieved from the $3/2_g$ spin level using control pulses and finally is re-emitted optically as photon echo at the frequency of the $1/2_g\leftrightarrow3/2_e$ transition.}
    \label{fig:setup}
\end{figure*}

\subsection{Entanglement generation scheme}

Our experimental setup, as depicted in Fig.~\ref{fig:setup}~(a), consists of two quantum nodes, A and B, each containing a non-degenerate cSPDC entangled photon-pair source and a solid-state quantum memory. the QMs are placed in separate cryostats located \SI{3}{\meter} apart. Signal photons are stored in the QM while idlers fields are mixed on a 50/50 beamsplitter (BS) at a central station. In the absence of losses in the signal modes and neglecting multiphoton emission, a heralding event projects the state of the QMs into a single-photon entangled state, where an atomic excitation is delocalized between the QMs of the two remote nodes $A$ and $B$ \cite{Duan2001,Lago-Rivera2021}:
\begin{equation}\label{eq:bellstate}
    \ket{\Psi^\pm} = \left( \ket{10}_{AB} \pm e^{i\Delta\varphi}\ket{01}_{AB} \right)/\sqrt{2},
\end{equation}
where $\ket{0},\ket{1}$ correspond to the state of a memory with 0 or 1 excitation stored, respectively, and $\Delta\varphi = \Delta\varphi_s + \Delta\varphi_i$ where $\Delta\varphi_s, \Delta\varphi_i$ denote the relative optical phase difference picked up by the signal and idler photons after they have been emitted by the source. The relative phase between the two components of the quantum state depends on which detector is used for heralding: a detection at idler detector $i_1$  or $i_2$ will herald the state $\ket{\Psi^+}$ or $\ket{\Psi^-}$, respectively.
We emphasize that the entanglement is heralded upon detection of a single idler photon and therefore may eventually offer higher entanglement rates than schemes relying on two-photon detections.

In order to consistently herald the same quantum states for each idler detection event, the phase term $e^{i\Delta\varphi} = e^{i(\Delta\varphi_s + \Delta\varphi_i)}$, has to be kept constant. To this end, we insert piezoelectric fiber stretchers (PZ) in signal and idler paths in order to stabilize the fiber length by locking to a classical interference signal (see Supplementary Material).

\subsection{On-demand quantum memory}

The quantum memory used in this work is a \ce{Pr^3+}:\ce{Y_2SiO_5} rare-earth doped crystal cooled down to \SI{3}{\kelvin} in a liquid helium closed-cycle cryostat (Montana Cryostation at node A and MyCryoFirm at node B). We exploit the broad inhomogeneous linewidth ($\approx\SI{10}{\giga\hertz}$) of the the $^{3}H_4(0) \leftrightarrow ^{1}D_2(0)$ optical transition at \SI{606}{\nano\metre} to perform light storage using the atomic frequency comb (AFC) protocol \cite{Afzelius2009}. The relevant energy levels are shown in Fig.~\ref{fig:setup}~(d) and the three states forming the lambda system used in the AFC storage sequence are denoted as $\ket{g},\ket{s}$ and $\ket{e}$. The memory is prepared by optically pumping the atomic ground state population such as to keep the hyperfine level $\ket{s}$ empty then by preparing an AFC of periodicity $\Delta$ in the $\ket{g}\leftrightarrow\ket{e}$ transition (see Supplementary Materials for details). As an incident photon gets absorbed,
a collective excitation is created and the photon is re-emitted at a time $\tau_{\text{AFC}}=1/\Delta$.
Since the parameter $\Delta$ has to be set in advance at the memory preparation stage, before an incoming photon enters the crystal, storage in the excited state using the AFC alone does not allow for on-demand retrieval of the stored excitation.

We implement on-demand retrieval by resorting to the complete spin-wave AFC protocol~\cite{Afzelius2010a}. After the creation of an optical excitation in the crystal, the excited state population $\ket{e}$ is transferred down to the long-lived hyperfine spin state $\ket{s}$, initially left empty during memory preparation. This is achieved by sending a bright optical control pulse (CP) resonant with the $\ket{s}\leftrightarrow\ket{e}$ transition that will freeze the AFC rephasing. After a time $T_s$, the population in $\ket{s}$ is mapped back onto the excited state $\ket{e}$ by a second CP, the AFC rephasing resumes and a photon is re-emitted at a time $\tau_{\text{AFC}}+T_S$. By tuning the delay $T_S$ between the two control pulses, the total storage time can be adjusted at will.
However, the control pulses are the main source of noise in this protocol as they can cause fluorescence stemming from a nonzero population remaining in $\ket{s}$ due to imperfect optical pumping, scattering of uncorrelated light into the photon collection mode, as well as spurious free-induction decay. To mitigate the effect of this noise, each QM is followed by an additional \ce{Pr^3+}:\ce{Y_2SiO_5} filter crystal featuring a narrow transparency window around the signal photon frequency (See Appendix A for details).

\subsection{Narrowband entangled photon-pair sources}

The cSPDC source used to generate entangled photon-pairs is designed such as to fulfill two main requirements, namely compatibility with telecom fiber links and compliance with storage in \ce{Pr^3+}:\ce{Y_2SiO_5} quantum memories \cite{Fekete2013}.
To this end, we use non-degenerate quasi-phase matching in a periodically-poled lithium niobate where a \SI{426}{\nano\metre} pump field generates a telecom idler photon at \SI{1436}{\nano\metre}, used for heralding over a fiber channel, and a signal photon at \SI{606}{\nano\metre}, which is stored in the QM.
Moreover, by embedding the nonlinear crystal into a doubly-resonant optical cavity, we ensure that the signal photons are generated in a \SI{2.5}{\mega\hertz} narrow-linewidth cavity mode that falls within the \SI{4}{\mega\hertz} bandwidth of AFC storage. To ensure that the heralded signal photons belong to a single well-defined frequency mode, we restrict the idler photon spectrum to a single mode using a narrow filter cavity, as depicted in Fig.~\ref{fig:setup}~(a).
The sources of node A and B feature a heralding efficiency of around \SI{20}{\percent} (in fiber), and detected raw idler count rate of around \SI{5300}{\cps} for a typical pump power of \SI{3.85}{\milli\watt}. To guarantee both spectral matching of the signal photons to the AFC and indistinguishability of the idler fields from both nodes a similar cascaded locking scheme of the two source cavities as in Ref.~\cite{Lago-Rivera2021} is implemented.

\section{Quantum Light Storage Performance}
\label{sec:g2}
\begin{figure}
    \centering
    \includegraphics[width=0.80\linewidth]{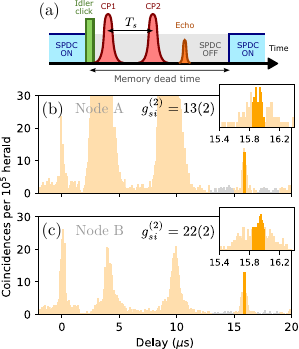}
    \caption{(a) Schematic representation of the spin-wave AFC storage sequence in conditional operation. Upon detection of an idler photon the pump of the cSPDC source is turned off and a pair of control pulses is sent into the memory with an arbitrary temporal separation $T_S$. A photon echo is then re-emitted after a time $\tau_{\text{AFC}}+T_S$. The memory remains closed for a fixed period of time (memory dead time) to allow for the CP2 fluorescence noise to decay before the next storage trial. The SPDC pump is also being turned off for a fixed time of \SI{20}{\micro\second}  (see main text for details). (b-c) Time-correlation histograms between detection of an idler photon and a signal photon retrieved from the on-demand spin-wave AFC QM at (b) Node A and (c) Node B. The dark orange region highlights the selected \SI{280}{\nano\second}-long detection window. Grey areas indicate the noise windows. The errors on the value of $g^{(2)}_{si}$ are given assuming poissonian statistics. Integration times are of 5 minutes and 15 minutes respectively for nodes A and B.} 
    \label{fig:g2s}
\end{figure}

We assess the quality of single photon storage independently at each node by measuring the temporal cross-correlation $g^{(2)}_{si}$ between an idler photon and a signal photon retrieved after spin-wave AFC storage. For this characterization, the memories operate at a total storage time of $\tau=\tau_{\text{AFC}}+T_S=\SI{16.5}{\micro\second}$ with $\tau_{\text{AFC}}=\SI{10}{\micro\second}$ and $T_S=\SI{6.5}{\micro\second}$. 
The experiment runs in a conditional fashion, as depicted in \cref{fig:g2s}~(a). 
Upon detection of an idler photon, a first CP is sent to the QM while the SPDC pump is turned off in order to avoid noise generated by the pump field from leaking inside the spin-wave echo temporal mode. Consecutive to the memory readout executed with a second CP, the SPDC turns back on after a total delay of \SI{20}{\micro\second} while the memory remains closed for a dead time of \SI{200}{\micro\second} during which no storage is performed in order to allow for fluorescence from CP2 to decay before the next storage attempt is performed. 
The pump off time needs to be longer than the storage time in the QM but shorter than the memory dead time. We choose to use the shortest value possible (\SI{20}{\micro\second}) to be able to reduce the dead time in measurements where a larger storage duty cycle is required (see Section VI and Appendix D).
During this memory dead time, idler counts are still being recorded but no CPs are sent upon detection, therefore effectively closing the memory.
Idler and signal photons are detected using superconducting nanowire single photon detectors (IDQuantique ID281) and the detection timestamps are processed using a time-tagger (Zurich Instruments HDAWG). 

The obtained time-correlation histograms are presented in~\cref{fig:g2s}~(b-c). 
We record storage efficiencies of \SI{3.25}{\percent} and \SI{4.50}{\percent} at Node A and Node B respectively with a \SI{280}{\nano\second}-wide detection window (see Appendix~\ref{sec:memcharac} for more details on memory characterization). This detection window contains around \SI{60}{\percent} of the total temporal distribution of the retrieved photon (see Appendix G for a discussion on the choice of detection window size).
We obtain the value of the cross correlation as $g^{(2)}_{si} = p_{si}/(p_s p_i)$ where $p_{si}$ is the probability of detecting a coincidence between idler and signal detectors within the detection window while $p_i$ and $p_s$ are the probabilities of signal and idler clicks. We experimentally estimate the product $p_sp_i$ by averaging the number of coincidence events over two noise windows located before and after the photon detection window (grey areas in \cref{fig:g2s}~(b-c)) which gives us an estimate of the noise baseline at the location of the echo. For a \SI{280}{\nano\metre}-wide detection window we obtain values of $g^{(2)}_{si,A}= \num{13\pm2}$ and $g^{(2)}_{si,B}= \num{22\pm2}$ which are, to our knowledge the highest numbers reported for single photon storage in a rare-earth based solid-state spin-wave quantum memory~\cite{Rakonjac2021}.

\section{Entanglement Verification Under Conditional Storage} \label{sec:entanglement}
\begin{figure*}
    \centering
    \includegraphics[width=0.75\linewidth]{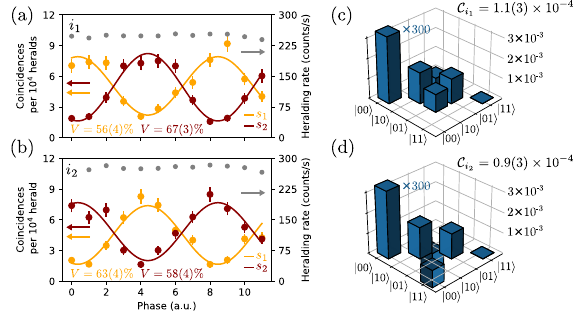}
    \caption{Verification of entanglement between spin-wave quantum memories. (a-b) Single photon interference of the signal modes at the memory output upon heralding from (a) detector $i_1$ and (b) detector $i_2$. Yellow (dark red) curve denotes the coincidences recorded with signal detector $s_1$ ($s_2$) and integration time is \SI{10}{\minute} per point. As expected from the unitarity of BS operation, we observe a $\pi$ phase shift between the interference measured at a given signal detector when considering heralding from one of the other idler detector. (c-d) Reconstructed density matrix for the quantum state heralded by (c) detector $i_1$ and (d) detector $i_2$. The total storage time is \SI{16.5}{\micro\second}.}
    \label{fig:tomo}
\end{figure*}

Taking advantage of the demonstrated high signal-to-noise ratio of on-demand storage in the QMs, we next demonstrate that an entangled state between spin-waves hosted by the two crystals is successfully heralded upon detection of an idler click at the central station. We follow the same procedure as described above in \cref{fig:g2s}~(a) and send control pulses upon idler detection.
We describe the heralded state of the photonic modes at the output of the two memories using the following density matrix~\cite{Chou2005,Lago-Rivera2021}:
\begin{equation}\label{eq:rho}
    \rho = 
    \begin{pmatrix}
        p_{00} & 0 & 0 & 0 \\
        0 & p_{10} & d & 0 \\
        0 & d & p_{01} & 0 \\
        0 & 0 & 0 & p_{11} \\
    \end{pmatrix},
\end{equation}
where $p_{ij}$ denotes the conditional probability to detect $i$ and $j$ photons at the output of memory $A$ or $B$, respectively, given a heralding event has been recorded. The diagonal and off-diagonal elements are estimated in two separate experimental runs, with dedicated setups depicted in~\cref{fig:setup}~(b-c).

The diagonal elements of the density matrix are measured in the configuration represented in Fig.~\ref{fig:setup}~(b), where timestamps of direct detection at each of the four detectors are being recorded. A single idler click not correlated to a detection after either of the two memory contributes to $p_{00}$, coincidences between an idler detector and signal detector $s_1$ (resp. $s_2$) add to $p_{10}$ (resp. $p_{01}$) while triple-coincidence events correspond to the $p_{11}$ term. 
The terms $p_{00}$, caused mainly by signal photon losses between generation and detection, and $p_{11}$, associated with spurious noise photon detection at the memory output, should approach zero in order to obtain a maximally entangled state.
Using the same storage time of $\tau=\SI{16.5}{\micro\second}$ as previously, we acquire timestamps for a total of \SI{2}{\hour}. Owing to the combination of low noise after the memory, reflected by the high measured values of $g^{(2)}_{si}$, and moderate transmission in the signal channel three-fold coincidence events are highly unlikely, which prevents direct measurement of $p_{11}$. Instead we estimate the value of this term from the measurement of $p_{10},p_{01}$ using the model described in Appendix~\ref{sec:p11_model}.
Experimental values for the diagonal elements are given in Table~\ref{tb:diags}. 
As a first step towards certifying entanglement, we compute the two-photon suppression $h^{(2)}_c = p_{11}/(p_{10}p_{01})$, analogous to a heralded autocorrelation. This quantity quantifies the relative amount of triple coincidence events arising from double excitations in the QMs but also detector dark counts and noise from the storage in the spin state. It is an indicator of the extent to which the system operates in a regime where a single photon delocalized over the two signal output modes is retrieved from the QMs. We measure experimentally a value of $h^{(2)}_c = \num{0.27\pm 0.01}$ smaller than 1, which is a necessary condition for the certification of entanglement in the photon-number basis~\cite{Chou2005}.
We stress that, using AFC storage, values of $h^{(2)}_c$ as low as \num{0.036} were demonstrated in a similar system~\cite{Lago-Rivera2021}, showing that this quantity is not fundamentally limited by the amount of double excitations present in the QMs but is rather dominated by the noise generated at the storage level by the CPs.

\begin{table}[h!]
\centering
 \begin{tabular}{|c | c |} 
 \hline
 Matrix element & Value \\ \hline
 $p_{00}$ & \num{0.999056 +- 0.000016}  \\  \hline
 $p_{10}$ & \num{4.4 \pm 0.1d-4}   \\ \hline
 $p_{01}$ & \num{5.0 \pm 0.1d-4}  \\ \hline
 $p_{11}$ & \num{5.9 \pm 0.1d-8}   \\ \hline
 $d_{i_1}$ & \num{3.0 \pm 0.1d-4} \\ \hline
 $d_{i_2}$ & \num{-2.9\pm 0.1d-4} \\ \hline
 \end{tabular}
 \caption{Experimentally measured matrix elements of $\rho$ for a storage time of \SI{16.5}{\micro\second}. $p_{10},p_{01}$ have been measured directly while $p_{11},p_{00}$ have been inferred from the model presented in Appendix F. The off-diagonal element $d$ is specified for the reconstructed state heralded by either idler detector $i_1$ or $i_2$.}
 \label{tb:diags}
 \end{table}

To complete the state tomography, we experimentally estimate the off-diagonal terms of $\rho$ as \cite{Chou2005}: $d=V(p_{10}+p_{01})/2$, where $V$ is the visibility of the single-photon interference obtained by mixing the output signal modes from both memories on a 50/50 BS, as depicted in Fig.\ref{fig:setup}~(c). To do so, we stabilize the phase of both idler and signal paths using piezo actuators in a feedback loop. We then shift the signal phase by applying a controlled bias voltage to the piezo in the signal path while recording twofold coincidence counts between the four configurations of idler-signal detector pairs (see Appendix C for details on the locking scheme). We use an integration time of \SI{10}{\minute} per point.
The number of coincidence counts for each of the four pairs of detectors as a function of the phase offset is plotted in \cref{fig:tomo}~(a-b). We highlight that there is a $\pi$ phase difference between the interference curves recorded by heralding on idler mode $i_1$ and $i_2$, arising from the unitarity of the beam-splitter operation. As a result, the two idler detectors effectively herald each one a different quantum state. 
From the extracted average visibilities $V_{i_1} = \SI{63\pm 3}{\percent}$ and $V_{i_2}=\SI{61 \pm 3}{\percent}$ we infer the values of the off-diagonal elements $d_{i1}$ and $d_{i2}$ reported in \cref{tb:diags}. 
During this measurement, the heralding rate remains constant with an average value of $\SI{249}{\cps}$ and $\SI{277}{\cps}$ for detectors $i_1$ and $i_2$ respectively. 
The single-photon visibility is limited by several mechanisms including background noise in the detection window, phase noise across the signal interferometer and imperfect signal and idler modal indistinguishability (see Appendix~\ref{sec:vis_est}). 
We also emphasize that in the present experiment, the combined heralding rate remains limited by the available pump power of the cSPDC source although already approaching the maximum of \SI{2.5}{\kilo\cps} set by memory preparation duty cycle and dead time (see Appendix~\ref{sec:swg2model}).

The reconstructed density matrices of the state heralded by each idler detector are displayed in \cref{fig:tomo}~(c-d). 
As entanglement witness, we use the concurrence $\mathcal{C}$, which is positive in the case of an entangled state.
It can be evaluated as $\mathcal{C} = \max(0,2d-2\sqrt{p_{00}p_{11}})$ \cite{Chou2005} and we obtain experimentally $\mathcal{C}_{i_1} = \num{1.1 \pm 0.3 d-4}$ and $\mathcal{C}_{i_2} = \num{0.9 \pm 0.3 d-4}$ at the detectors thus demonstrating positive concurrence by respectively \num{3.75} and \num{3.25} standard deviations of the state heralded by detector $i_1$ and $i_2$. The low absolute value of $\mathcal{C}$ can mostly be attributed to the dominant contribution from $p_{00}$ to the heralded states, which in turn can be attributed to the limited heralding efficiency of the cSPDC sources as well as losses in the signal channel. To set aside the impact of losses on the value of $\mathcal{C}$ we compute the concurrence back-propagated to the inside of the crystals $\Tilde{\mathcal{C}}$ by correcting for the signal detector efficiency $\eta_\text{det} = \SI{71}{\percent}$, the setup losses $\eta_\text{setup} = \SI{10}{\percent}$ and the readout efficiency of the memories $\eta_\text{read}^{(A)} = \SI{6.25}{\percent}$, $\eta_\text{read}^{(B)} = \SI{8.65}{\percent}$ (see Appendices~\ref{sec:memcharac} and~~\ref{sec:setupcharac} for details). We indeed obtain values two orders of magnitude higher with $\Tilde{\mathcal{C}}_{i_1} = \num{3.5 +- 0.6 d-2}, \Tilde{\mathcal{C}}_{i_2} = \num{2.9 +- 0.6 d-2}$ with violations of \num{6.2} and \num{5.2} standard deviations respectively for idler detectors $i_1$ and $i_2$.

However, in view of the application of our system in a Duan-Lukin-Cirac-Zoller (DLCZ)-like quantum repeater, this vacuum contribution can be discarded thanks to the inherent built-in entanglement purification provided by the protocol~\cite{Duan2001}. Indeed, in such an architecture 
two-photon polarization-like entanglement is generated upon two subsequent heralding events, but it is also followed by post-selection on simultaneous signal photon detection between opposite nodes \cite{Chou2007}, thus effectively removing any contribution from vacuum terms in the distributed quantum state. 
Consequently, we estimate the performance of our system in the presence of post-selection using as a metric the effective fidelity $\mathcal{F}_\text{eff}$ to the maximally entangled state of \cref{eq:bellstate} when restricting the density matrix to the subspace spanned by single and two photon Fock-states. We obtain a value of $\mathcal{F}_{\text{eff}} = \SI{81\pm 1}{\percent}$ for states heralded by both detectors $i_1$ and $i_2$, well above the \SI{50}{\percent} required for certifying entanglement. We measure an associated detection rate of \SI{0.32}{\cps} that is  given by the sum of the detected twofold coincidences per unit time recorded over all four configurations of idler-signal detector pairs and corrected for the losses of the signal mixing station.
We point out that the difference between the value of the detection rate of and of the heralding rate is not fundamentally limited and is simply due to the non-unity heralding efficiency of the source, memory storage efficiency, setup transmission and detector efficiency. Consequently, by increasing these quantities the detection rate should approach the heralding rate.

Lastly, we stress that the demonstrated value for the heralding rate is only limited by the available source pump power. Indeed, the cross-correlation values obtained by storing in an AFC only without CPs remain high, \num{92(19)} for Node A and \num{157(32)} for Node B. Therefore, as shown in Appendix~\ref{sec:swg2model}, we could afford to use a \num{3.9} times larger pump power per source while still maintaining $g^{(2)}_{si,A},g^{(2)}_{si,B}$ high enough to observe entanglement.

\section{Quantum State Manipulation Using Active Feedforward} \label{sec:ff}
\begin{figure*}
    \centering
    \includegraphics[width=\linewidth]{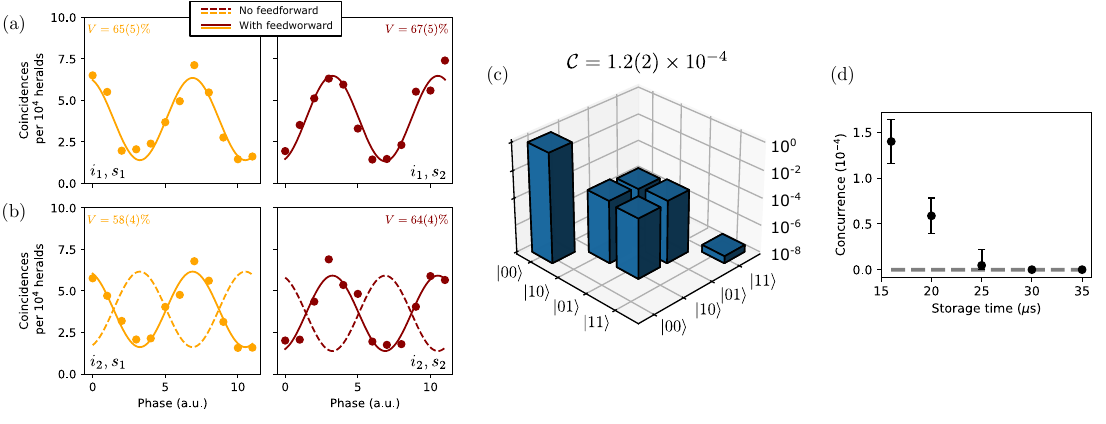}
    \caption{Dynamical phase feed-forward using conditional readout of the spin-wave memories. Single photon interference at the signal 50/50 BS upon heralding at (a) idler detector $i_1$ and (b) idler detector $i_2$. Solid lines represent a sinusoidal fit of the data. Dashed lines correspond to the expected unshifted interference fringe in the absence of conditional phase feed-forward. We observe that, as a result of the feed-forward, interference fringes are identical between $i_1$ and $i_2$ indicating that both idler detectors now herald the same quantum state. The integration time is \SI{10}{\minute} per point. (c) Reconstructed density matrix of the state of photonic modes at the output of the QMs including phase-corrected heralding from both idler detectors. The total storage time is \SI{16.5}{\micro\second}. (d) Evolution of measured concurrence with total storage time in the spin wave quantum memories. The grey dashed line denotes the $\mathcal{C}=0$ threshold for the state to be entangled.}
    \label{fig:ff}
\end{figure*}

After establishing experimental evidence for entanglement between the memories, we further demonstrate the versatility of our scheme by exploiting the possibility to control the phase of the retrieved photon directly at the read-out stage.
Indeed, by adjusting the phase of the RF waveform used to generate the second CP a corresponding shift is imprinted on the optical phase of the read-out photon.
This not only provides us with an elegant way to perform single-photon interference without physically acting on the interferometer itself but also gives us a mean to dynamically correct for the $\pi$ phase difference caused from heralding at different idler detectors.
In what follows, we dynamically feed back the information of which idler detector recorded the heralding signal in the form of a conditional $\pi$ phase shift applied to the second control pulse of memory at Node A. As a result, a detection event at any of the two idler detectors will indistinctly herald the exact same quantum state at the output of the memories. This technique is experimentally demonstrated in~\cref{fig:ff}~(a-b) where we show the obtained single-photon interference curves after the signal 50/50 BS by scanning the optical phase using the CPs and applying dynamical phase feed-forward. We observe that the $\pi$ phase difference between the curves displayed \cref{fig:ff}~(a) and (b) is compensated, in contrast to the data of \cref{fig:tomo}~(a-b). Besides we notice no decrease in the interference visibility, with measured values of $V_{i_1} = \SI{65\pm 3}{\percent}$ and $V_{i_2}=\SI{61\pm3}{\percent}$ while effectively doubling the heralding rate to \SI{510}{\cps} for a given quantum state. The reconstructed density matrix under phase feedforward is plotted in~\cref{fig:ff}~(c). Thanks to the larger count rates, the error on the measured concurrence at the detectors can be further reduced, reaching a positive value by 6.45 standard deviation with $\mathcal{C} = \num{1.2 \pm 0.2 d-4}$. This measurement shows the outstanding potential of our setup to distribute entanglement at high rate on a local area scale (up to 2 km) where communication times are shorter than the storage time in the excited state (10 µs). 

Taking advantage of this improved heralding rate, we give a direct experimental proof of the on-demand character of photon storage by varying the time $T_S$ and repeating quantum state tomography to assess the degree of entanglement at different storage times. The measured value of the concurrence at the detectors as a function of total storage time is shown in~\cref{fig:ff}~(d). We see that $\mathcal{C}$ remains greater than zero for more than \SI{20}{\micro\second} storage time, corresponding to an equivalent fiber link length of around \SI{4}{\kilo\metre}. The main mechanism contributing to the loss of entanglement is the decrease of memory efficiency with storage time and the subsequent decrease of signal-to-noise ratio caused by dephasing of the spin excitation due to the inhomogeneous broadening of the $\ket{g}\leftrightarrow\ket{s}$ hyperfine transition (see Appendix A for details). Nevertheless, this measurement shows that thanks to the robustness of the generated entanglement and its compliance with telecom fiber links, our scheme could be deployed at a metropolitan scale.

\section{Temporal Mulptiplexing with Unconditional Storage} \label{sec:mux}
\begin{figure}
    \centering
    \includegraphics[width=0.85\linewidth]{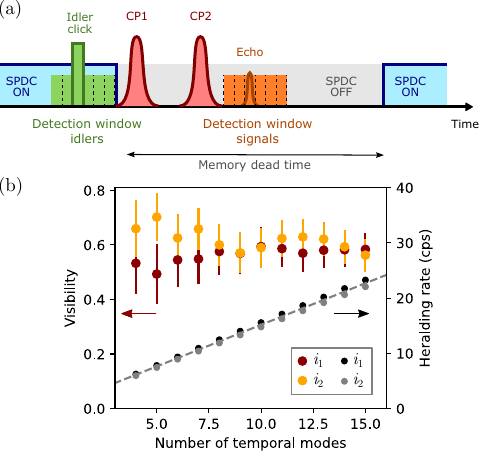}
    \caption{Temporally-multiplexed entanglement generation. (a) Sketch of the unconditional storage sequence used in this measurement. The SPDC pump is turned on and off and CPs are being sent at a fixed repetition rate, defining a memory acceptance window of 15 temporal modes. (b) Single-photon visibility at the signal 50/50 BS with heralding on both $i_1$ and $i_2$ detectors (left y-axis) and corresponding heralding rates (right y-axis) as a function of the number of stored temporal modes. The total storage time is \SI{16.5}{\micro\second} and the integration time is \SI{1}{\hour} per point. The dashed line is a linear fit of the average heralding rate with the number of modes.}
    \label{fig:mux}
\end{figure}

Lastly, we showcase the inherent temporal multimode capacity of the spin-wave AFC protocol and demonstrate how it can be exploited to increase the heralding rate without compromising the degree of entanglement. To this end, we run the experiment in an unconditional fashion to simulate the communication time of long fiber links, as sketched in \cref{fig:mux}~(a). Contrary to the conditional operation used in Section~\ref{sec:entanglement}, here the delivery of the first CP at the QMs and the switching off of the SPDC pump are not carried out upon detection of an idler photon at the central station. Instead, the SPDC pump is periodically switched on and off at a rate given by the memory dead time, which was set to \SI{100}{\micro\second} for this measurement. After switching on, the memory is open for a fixed duration of \SI{6}{\micro\second} corresponding to 15 temporal modes of \SI{400}{\nano\second}, during which idler counts are recorded before the first CP is being sent, thus effectively ``closing'' the memory acceptance window. The SPDC is then also turned off.
After the second CP, signal photons exit the memory in a temporal mode that matches the temporal mode of the idler.
In this scheme the temporal multimodality of the spin-wave AFC protocol allows for photons arriving at the memory at any temporal mode during the memory acceptance window to be stored and retrieved on-demand. In contrast, in a QM exempt of temporal multiplexing the memory acceptance window only features a single temporal mode thus reducing drastically the duty cycle of the storage sequence. 

To exemplify the increase in heralding rate thanks to temporal multiplexing, we record single photon interference of the signal photons (setup of \cref{fig:setup}~(c)) while running the storage sequence unconditionally. The total storage time is \SI{16.5}{\micro\second} and phase correction by feedforward is not being applied. In post-processing, we restrict the timestamps that are used to reconstruct the interference fringes to an increasing number of temporal modes within the memory acceptance window. In \cref{fig:mux}~(b), we plot the values of the heralding rates and single-photon interference visibilities for both idler detectors $i_1,i_2$ as a function of the number of temporal modes included in the timestamps analysis, thus reflecting the effect of increasing the number of stored temporal modes in the memory. We witness a linear increase of the heralding rate as a function of the number of modes with a fitted slope of \num{1.505(1)} cps/mode and a maximum value of \SI{22}{\cps} per idler detector, limited by the available pump power at the sources. Importantly, this increase in rate is obtained without noticeable loss of entanglement, which can be seen from the fact that the single-photon visibilities remain essentially constant over the considered range of temporal modes. We stress that, although the heralding rates are reduced due to a shorter duty cycle under unconditional storage, this configuration is representative of the true operation of our system over a network. Indeed, the spin-wave storage process cannot be conditioned on the detection of an idler due to the fact that the incompressible latency of the idler detection signal when operating over long-distance fiber links is longer than $\tau_{AFC}$ for distances above a few kms.

\section{Discussion and Outlook}
In this work, we presented the demonstration of multiplexed heralded entanglement of on-demand ensemble-based quantum memories, an essential building block for future quantum networks. The low level of noise in on-demand storage of single photons allows for the measurement of entanglement for storage times of more than \SI{20}{\micro\second}. Dynamical control of the heralded quantum state through adjustment of the CP phase has been showcased for the first time in rare-earth doped crystal QMs, allowing for a two-fold increase in heralding rates for a specific state, up to 510 cps, with corresponding detection rate of \SI{0.32}{\cps}. When operated in conditional fashion, these results show that our scheme allows fast entanglement distribution over local area networks (few km). Lastly, we operated the system unconditionally, emulating metropolitan distance operation and showed how temporal multiplexing in the spin-wave AFC protocol results in a linear increase of the heralding rate with the number of modes. Altogether, the demonstrated results establish our system as a solid candidate for the implementation of scalable long-distance quantum networks.

In addition to providing a milestone for future developments toward DLCZ-like quantum links, our work opens the way to several extensions and improvements.
A first important step forward can be taken by deploying our scheme over a city-scale fiber telecommunication channel. This would require local entanglement verification using homodyne measurements~\cite{Caspar2020} as well as active correction of phase-noise in the idler optical fiber. The latter, although technically challenging, has already been implemented in several deployed quantum networks testbeds~\cite{Bersin2024, Liu2024, Stolk2024}. 
Moreover, we point out that, provided the idler phase lock remains stable and the number of heralding events is above dark-count level, the addition of extra fiber length will only be detrimental to the heralding rates without affecting the measured concurrence.
As discussed previously, the heralding rates are currently limited by the available pump power and could be increased by a factor of \num{3.9} using an optimized \SI{426}{\nano\metre} laser setup (see Appendix \ref{sec:swg2model}).
In this configuration, for an elementary repeater segment consisting of two metropolitan fiber links of \SI{10}{\kilo\metre} each displaying around \SI{2}{\decibel} losses, the total heralding rate of our single-click protocol in unconditional operation with a \SI{100}{\micro\second} memory dead time would be \SI{111}{\cps} at a total distance of \SI{20}{\kilo\metre}, leading to a detection rate of \SI{0.071}{\cps} with current values of memory efficiency and setup transmission, which compares favorably to recent realizations in telecom-heralded long distance experiments with similar fiber-link distance (see Table~\ref{tb:comparison} in the Appendix).

In order to reach their full potential for quantum communication nodes, our spin-wave AFC QMs will be upgraded in future work with several features that have already been demonstrated in separate experiments.
High storage efficiency can be achieved by embedding the rare-earth doped crystal in an impedance-matched cavity \cite{Duranti2024} while storage time in the spin state is expected to reach the 10 millisecond regime with negligible drop in efficiencies at shorter timescales (around \SI{250}{\micro\second}) using dynamical decoupling in combination with a weak static magnetic field \cite{Etesse2021, Ortu2022a, Pignol2024} or even tens of seconds by operating in a zero first order Zeeman (ZEFOZ) configuration~\cite{Hain2022}. 
In addition, the spin-wave AFC protocol supports parallel multiplexing in more than one degree of freedom \cite{Ortu2022,Sinclair2014,Seri2019,Yang2018,Teller2025} thus allowing to scale up even further the heralding rate. Rare-earth doped QM have also been proven to be compatible with chip-integration, providing an avenue for portable and cost-effective quantum nodes~\cite{Zhong2017, Rakonjac2022, Zhu2022, Zhou2023}.

Lastly, a natural continuation to the present work consists in the implementation of a full DLCZ-like repeater link featuring two SPDC sources and four QMs~\cite{Duan2001,Chou2007,Simon2007}. In addition to providing a more natural two-photon qubit encoding, this scheme is expected to achieve high success rates approaching the heralding rate of a single source when using efficient and long-lived QM. This architecture is also well suited to long-distance scenarios thanks to less stringent phase stability requirements as well as a straightforward entanglement verification process.

\section*{Acknowledgements}
This project received funding from: Gordon and Betty Moore Foundation (GBMF7446 to H. d. R); Agència de Gestió d'Ajuts Universitaris i de Recerca; Centres de Recerca de Catalunya; Fundació Privada MIR-PUIG; Fundación Cellex; Ministerio de Ciencia e Innovación with funding from European Union NextGeneration funds (MCIN/AEI/10.13039/501100011033, PLEC2021-007669 QNetworks, PRTR-C17.I1); Agencia Estatal de Investigación (PID2023-147538OB-I00, Severo Ochoa CEX2019-000910-S); European Union research and innovation program within the Flagship on Quantum Technologies through Horizon Europe project QIA-Phase 1 under grant agreement no. 101102140 and from the Secretariat of Digital Policies of the Government of Catalonia - G.A. GOV/51/2022. F.A. and M.T. acknowledge funding from the European Union's Horizon 2022 research and innovation programme under the Marie Sklodowska-Curie grant agreements No 101104148 (IQARO) and 101103143 (2DMultiMems) respectively. S.G. acknowledges funding from ``la Caixa'' Foundation (ID 100010434; fellowship code LCF/BQ/PR23/11980044). J.H. acknowledges funding from the “Secretaria d’Universitats i Recerca del Departament de Recerca i Universitats de la Generalitat de Catalunya” under grant 2024 FI-2 00059, as well as the European Social Fund Plus.

\section{Appendix}
\begin{table*}
\centering
 \begin{tabular}{|c | c | c | c |} 
 \hline
 \textbf{Physical system} & \textbf{Reference} & \textbf{Link distance (km)} & \textbf{Detection rate (counts/s)} \\ \hline
 NV centers & \cite{Stolk2024} & \SI{25}{\kilo\meter} & 0.022 \\ \hline
 SiV centers & \cite{Knaut2024} & \SI{20}{\kilo\meter} & 0.0029 \\ \hline
 Single neutral atoms & \cite{vanLeent2022} & \SI{33}{\kilo\meter} & 0.012 \\ \hline
 Cold atomic clouds & \cite{Liu2024} & \SI{20}{\kilo\meter} & 0.14 \\\hline
 Rare-earth doped crystals & This work (estimated) & \SI{20}{\kilo\meter} & 0.071 \\\hline
 \end{tabular}
 \caption{Comparison of detection rates for telecom-heralded entanglement between remote quantum nodes across different physical platforms. The performance for the present work is estimated in unconditional operation assuming a 4-fold increase in SPDC pump power, as detailed in Appendix~\ref{sec:swg2model}.}
 \label{tb:comparison}
 \end{table*}

\subsection{Characterization of the quantum memories}
\label{sec:memcharac}
\begin{figure}
    \centering
    \includegraphics[width=\linewidth]{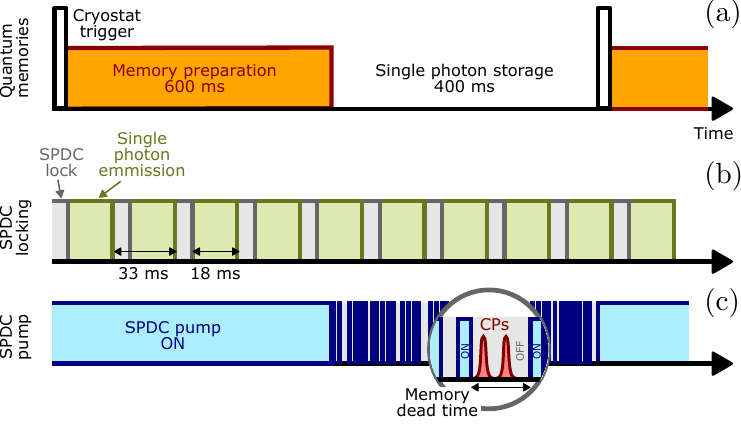}
    \caption{Schematic of the experimental sequence of the quantum memories and SPDC sources. The different durations are not to scale.}
    \label{fig:sequence}
\end{figure}

The quantum memories used in this work are \ce{Pr^3+}:\ce{Y_2SiO_5} doped with a concentration of \SI{0.05}{\percent}. The crystals (dimensions 2x3x5 mm) are cut along the $(D_1,D_2,b)$ crystallographic axes and the input beam is propagating along the $b$-axis. The memory preparation is performed with a frequency doubled \SI{606}{\nano\metre} diode laser (Toptica TA-SHG Pro) locked to an external home-built reference cavity. To prepare the AFC \cite{Nilsson2004, Afzelius2010a}, first a \SI{18}{\mega\hertz} wide transparency window (spectral pit) is burnt within the $^{3}H_4(0) \leftrightarrow ^{1}D_2(0)$ line of \ce{Pr^3+}:\ce{Y_2SiO_5}. Then this pit is repopulated with ions from different classes using a \SI{4}{\mega\hertz}-chirped and \SI{28.1}{\mega\hertz}-detuned burn-back pulse. Using cleaning pulses of the same chirp but detuned by \SI{6}{\mega\hertz} from the pit center, a \SI{4.6}{\mega\hertz}-wide single class feature is obtained on the $1/2_g\leftrightarrow3/2_e$ transition. The AFC is then carved out of this single class feature by sending repeated pulses whose temporal profile is the Fourier transform of the target AFC spectrum~\cite{Jobez2016}. Square cleaning pulse resonant with the $3/2_g\leftrightarrow3/2_e$ pulses are inserted after each repetition of the AFC burning to keep the spin level $3/2_g$ empty for spin-wave storage. Finally a last series of extra cleaning pulses are sent to the memory before launching the first storage attempt. The filter crystal used to filter out noise from the CPs is prepared by burning a square spectral hole of \SI{3.4}{\mega\hertz} (\SI{2.6}{\mega\hertz}) around the AFC spectral frequency for node A (B). Finally, the control pulse exhibit a Gaussian waveform of duration \SI{5}{\micro\second} and chirp of \SI{2.6}{\mega\hertz}.

As shown in~\cref{fig:sequence}, the preparation of each memory takes around \SI{600}{\milli\second} and is coordinated with the cooling cycle of the cryostat (approximately \num{60} per minute). The latter causes inevitable vibrations that affect negatively the coherence properties of the crystal. As a result the memory preparation is performed cyclically after each cryostat beat in the time range where those vibrations have the smallest amplitude. Since the two memories are cooled down by different cryostats (Montana Cryostation at Node A and MyCryoFirm custom cryostat at Node B) with slightly different pulsing periods and yet need to be prepared simultaneously, the preparation of Node B follows the cycles of the cryostat of Node A. Nevertheless, the vibration isolation of QM B was found to be good enough to not notice any appreciable loss of performance when operating un-syncrhonized with respect to the cycles of its own host cryostat.

\begin{figure}
    \centering
    \includegraphics[width=\linewidth]{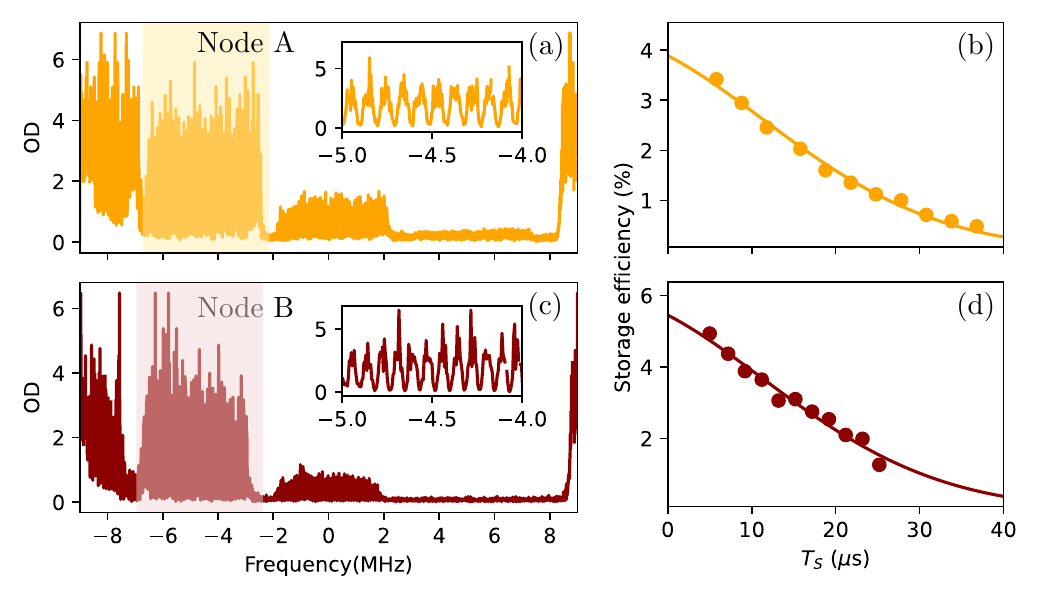}
    \caption{Quantum memory characterization. Measured absorption profile of the QM and spin-wave storage efficiency using weak coherent states for (a-b) Node A and (c-d) Node B. The highlighted regions in (a,c) correspond to the AFC and the insets on  show a \SI{1}{\mega\hertz} displaying the periodic structure of the absorption profile.}
    \label{fig:qm_charac}
\end{figure}

The experimentally-measured absorption profile of the quantum memories after preparation are displayed in~\cref{fig:qm_charac}~(a) and (c) where the AFC region is highlighted. In addition, we characterized the spin-wave storage efficiency using attenuated laser pulses as an input, with mean photon number of \num{0.85} and \num{0.83} for nodes A and B, respectively. After each memory preparation cycle, a total of 400 (node A) and 60 (node B) storage attempts are performed and the number of coincidence counts in a \SI{280}{\nano\second}-wide detection window are recorded. To estimate the number of input counts and compute the storage efficiency we measure the number of clicks in a similar \SI{280}{\nano\second}-wide window when photons are going through a transparency window prepared in the crystal and polarized in the non-interacting polarization of the material. The results are plotted in~\cref{fig:qm_charac}~(b,d). The decay of efficiency with respect to storage time is attributed to the dephasing imparted by the inhomogeneous broadening $\gamma_\text{inh}$ of the spin transition $\ket{g}\leftrightarrow\ket{s}$ which can be modeled by the following relation: $\eta_\text{sw} = \eta_0\eta_\text{inh} = \exp[-(T_S\gamma_\text{inh}\pi)^2/(2\ln{2})]$ with $T_S$ the time between the two control pulses during which the optical excitation is stored in the spin state $\ket{s}$. The solid lines in~\cref{fig:qm_charac}~(b,d) represent a fit of the data to this model from which we extract a value of $\gamma_\text{inh}=\SI{12.5}{\kilo\hertz}$ for both QMs.

To perform the backpropagation of the value of the concurrence inside the QMs in Section~\ref{sec:entanglement}, we estimated the readout efficiency of the memory as: 
\begin{equation}
    \eta_\text{read} = \frac{\eta_\text{sw}}{\eta_{write}\eta_\text{inh}}
\end{equation}
where $\eta_{write}=\SI{62.5}{\percent}$ is the spin-wave AFC write efficiency. The latter is calculated by assuming that $1-\eta_{write}$ corresponds to the ratio between the rate of coincidences of photons transmitted through the AFC spin-wave QM without being stored and the rate of coincidences of orthogonally-polarized photons transmitted through a transparency window. The spin dephasing factor $\eta_\text{inh}$ follows the expression defined in the previous paragraph. At a total storage time of $\tau=\tau_{\text{AFC}}+T_S=\SI{16.5}{\micro\second}$ with $\tau_{\text{AFC}}=\SI{10}{\micro\second}$ and $T_S=\SI{6.5}{\micro\second}$ we have $\eta_\text{sw}^{(A)} = \SI{3.25}{\percent}$, $\eta_\text{sw}^{(B)} = \SI{4.5}{\percent}$ and $\eta_\text{inh} = \SI{80.4}{\percent}$ which leads to $\eta_\text{read}^{(A)} = \SI{6.25}{\percent}$, $\eta_\text{read}^{(B)} = \SI{8.65}{\percent}$.

\subsection{Characterization of the setup transmission}
\label{sec:setupcharac}

The transmission of the setup from after the QMs to before the fiber leading to the SNSPDs (including going through the 606 nm mixing station) has been measured on a daily basis using a continuous-wave laser beam and has a value of \SI{20}{\percent}. To this value, we add a correction taking into account the reduced transmission of the narrow spectral pit of the filter crystal when using single photons which is estimated to be around \SI{50}{\percent}. This results in a value of $\eta_\text{setup} = \SI{10}{\percent}$ for the overall setup transmission. Finally,  the SNSPDs have a nominal detection efficiency of $\SI{80}{\percent}$ at $\SI{606}{\nano\meter}$ to which we need to incorporate the losses associated to two fiber-to-fiber connections of $\SI{94}{\percent}$ each, which corresponds to a combined detection efficiency of $\eta_\text{det}=\SI{71}{\percent}$.

\subsection{Stabilization of the cSPDC sources and of the idler and signal photon phases}

In this section, we describe the experimental procedure used in this experiment to lock the cSPDC sources and signal and idler phases.  
A nearly identical setup has been used and described in great details in ref.~\cite{Lago-Rivera2021}, where the reader can find additional details.
The SPDC sources are based on ppLN crystals that are temperature stabilized with a Peltier element. Both the pump laser at \SI{426}{\nano\meter} (Toptica DL Pro) and the preparation laser for the QM at \SI{606}{\nano\metre} (Toptica TA-SHG Pro) are locked to separate reference cavities using a standard Pound-Drever-Hall (PDH) technique. The pump powers used for this experiments are \SI{2.95}{\milli\watt} (\SI{3.65}{\milli\watt}) for node A (B). The SPDC sources operate in a lock-and-measure fashion with repetition rate of \SI{30}{\hertz} as depicted schematically in~\cref{fig:sequence}. The locking cycle of the sources is not synchronized with the memory preparation, the repetition rate of the latter being much lower ($\approx$ 1 /s).

In the first part of this cycle of duration $t_\text{lock} = \SI{10.5}{\milli\second}$, the SPDC sources are locked. A \SI{606}{\nano\meter} reference beam at the AFC frequency is sent to both source cavities. They are locked to be resonant with this light using a PDH loop feeding back on the cavity length, adjusted with a piezoelectric element mounted on one mirror. At the same time, the \SI{426}{\nm} pump laser frequency and \SI{606}{\nano\metre} light inside the cavity produce classical light through difference frequency generation (DFG) at the idler wavelength \SI{1436}{\nano\metre}. The pump frequency is locked in order to get resonant DFG emission inside one of the cavity modes at Node A.
This is done by frequency-modulating the pump beam using an acousto-optic modulator with a carrier frequency of \SI{110}{\mega\hertz} and modulation depth of \SI{150}{\kilo\hertz} at a rate of \SI{33}{\kilo\hertz} and feeding back the demodulated DFG signal onto the \SI{426}{\nano\metre} reference cavity length (controlled by a piezo on one of the mirrors). Heralding on a single idler frequency mode is ensured by filtering the generated \SI{1436}{\nano\metre} light with a free-space filter cavity.
During the entirety of this first time slot, mechanical choppers are used to prevent classical \SI{606}{\nano\metre} reference light and \SI{1436}{\nano\metre} DFG from damaging the SNSPDs.

In the second part of the locking cycle of duration $t_\text{meas}=\SI{18}{\milli\second}$ single photons are measured. The \SI{606}{\nano\meter} reference beam is blocked before the sources and all locks are on hold. Detectors are simultaneously unblocked and single photons can be detected.

When performing single-photon interference of signal fields for measurement of the coherences of $\rho$ (setup~\cref{fig:setup}~(c)), stabilization of the signal and idler phases is performed in the locking slot additionally to cavity locks. In this configuration, the duty cycle is modified as follows: i) $t_\text{lock}=\SI{3}{\milli\second}$ cavity lock (\SI{606}{\nano\metre} and DFG) + phase lock of idlers, ii) $t_\text{phase}=\SI{5}{\milli\second}$  phase lock of signal, iii) $t_\text{meas}=\SI{18}{\milli\second}$ single photon measurement. The idler phase is stabilized by looking at the classical interference of the DFG beams after the idler 50/50 BS with a side-of-the-fringe lock feeding back on the fiber stretcher shown in~\cref{fig:setup}~(a). For signal phase lock, the \SI{606}{\nano\meter} reference beam is transmitted through the source cavity then sent though the memory and interfered on the signal 50/50 BS. A similar side-of-the fringe lock as in the idler case is performed. To avoid affecting the memory preparation, the classical 606 light used to lock the phase of the signal interferometer is frequency shifted by \SI{320.6}{\mega\hertz} such that it is significantly detuned from the AFC central frequency.

\begin{figure}
    \centering
    \includegraphics[width=0.8\linewidth]{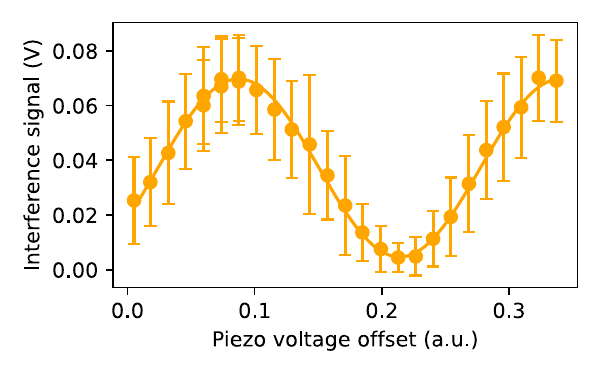}
    \caption{Classical lock-and-measure measurement of \SI{606}{\nano\metre} signal beam interference. The value of the visibility is extracted from the sinusoidal fit shown as a solid line. Error bars are calculated as the standard deviation of the averaged interference signal over each measurement window $t_\text{meas}$.}
    \label{fig:classvis}
\end{figure}

The signal phase locking is crucial since phase noise is important in fibers connecting source and memory optical tables. To assess the quality of the stabilization we perform a classical measurement of interference where we alternate between locking the phase for a time $t_\text{phase}$ then hold the lock for $t_\text{meas}$ after applying an increasing voltage offset to the piezoelectric fiber stretcher. The result is shown in \cref{fig:classvis} where each point corresponds to the detected interference for each voltage offset signal averaged over the detection time $t_\text{meas}$. We obtain a visibility of \SI{87+-1}{\percent} that is limited by mechanical stability of the fibers connecting the different parts of the interferometer.

\subsection{Effect of the memory dead time on heralding rate and concurrence}
\begin{figure}
    \centering
    \includegraphics[width=0.9\linewidth]{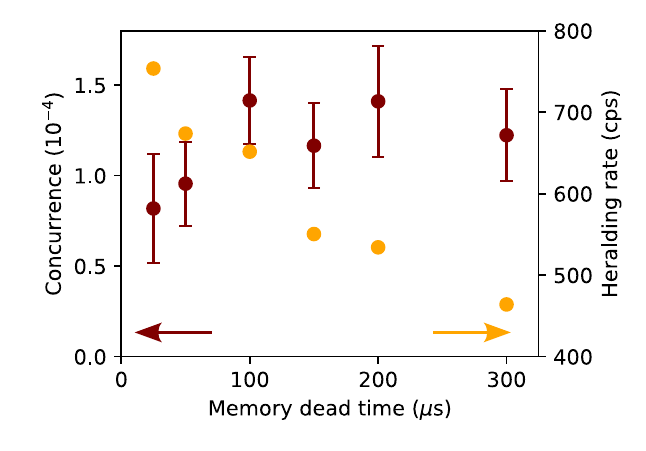}
    \caption{Measured concurrence and heralding rate as a function of the memory dead time. We observe that although the concurrence remains constant, the heralding rate can be significantly increased by lowering the memory dead time.}
    \label{fig:dtstudy}
\end{figure}

After each heralding event the memory is closed for a fixed duration which ultimately limits the achievable heralding rate. To assess the ultimate performance of the system, we characterize the heralding rate and concurrence for various memory dead times. The result is presented in~\cref{fig:dtstudy}. We see that the memory dead time can be made as short as \SI{25}{\micro\second} with positive concurrence. For this value, the heralding rate reaches \SI{750}{\cps} which represents a \SI{25}{\percent} increase with respect to the data presented in~\cref{fig:ff}. Nevertheless, we point out that in a realistic scenario the duty cycle of entanglement attempts will rather be set by the communication time of the idler photon through the fiber link (around \SI{250}{\micro\second} for a \SI{50}{\kilo\metre} distance).
As a consequence, for maximizing the rate at a given link distance, the memory dead time has to be chosen to match the idler communication time.
Through this auxiliary measurement, we therefore demonstrate that entanglement is preserved even when tuning this value, making our system compatible with a range of link distance from around \num{10} to \SI{60}{\kilo\metre}.

\subsection{Contributions to the single photon visibility}
\label{sec:vis_est}
The single photon visibility is affected by several sources of imperfections: signal and idler modal overlap at the beam-splitters $\eta_s,\eta_i$, phase noise in signal and idler optical path and, lastly, the signal-to-noise ratio of photon storage, reflected by the value of the cross-correlation $g^{(2)}_{si}$. The obtained visibility can be estimated as:
\begin{equation}\label{eq:vis_est}
    V = \eta_s\eta_iV_{606,\text{class}}V_{\text{DFG,class}}\dfrac{g^{(2)}_{si}-1}{g^{(2)}_{si}+1}.
\end{equation}

The modal indistinguishability of the idler fields has been obtained from a Hong-Ou-Mandel measurement and shows a value of $\eta_i = \SI{90+-7}{\percent}$ \cite{Lago-Rivera2021}. Although it has not been measured directly, we estimate the corresponding value $\eta_s$ for the signal mode to be of the same order as $\eta_i$. The DFG visibility $V_{\text{DFG,class}}$ is obtained by looking at the classical interference between the generated DFG light from the two sources during the locking stage and gives a value of $\SI{95}{\percent}$. The effect of phase noise in the DFG lock is typically very small since the idler optical fibers and beam-splitter are resting on the optical table and are therefore subject to little vibrations. The value of the classical visibility of the signal field $V_{606,\text{class}}$ is extracted from the measurement presented in~\cref{fig:classvis} with a reported value of $\SI{87+-1}{\percent}$. Lastly, the mean value of the cross-correlation obtained at each memory is $g^{(2)}_{si} = \num{17+-2}$ which gives $(g^{(2)}_{si}-1)/(g^{(2)}_{si}+1) = \SI{89+-1}{\percent}$. By including all these different coefficients, we arrive at an estimated value for the single-photon visibility of \SI{60(7)}{\percent} which is in good agreement with the numbers reported in~\cref{fig:tomo,fig:ff}.

\subsection{Model for $p_{11}$ estimation and its experimental validation}
\label{sec:p11_model}

\begin{figure}
    \centering
    \includegraphics[width=\linewidth]{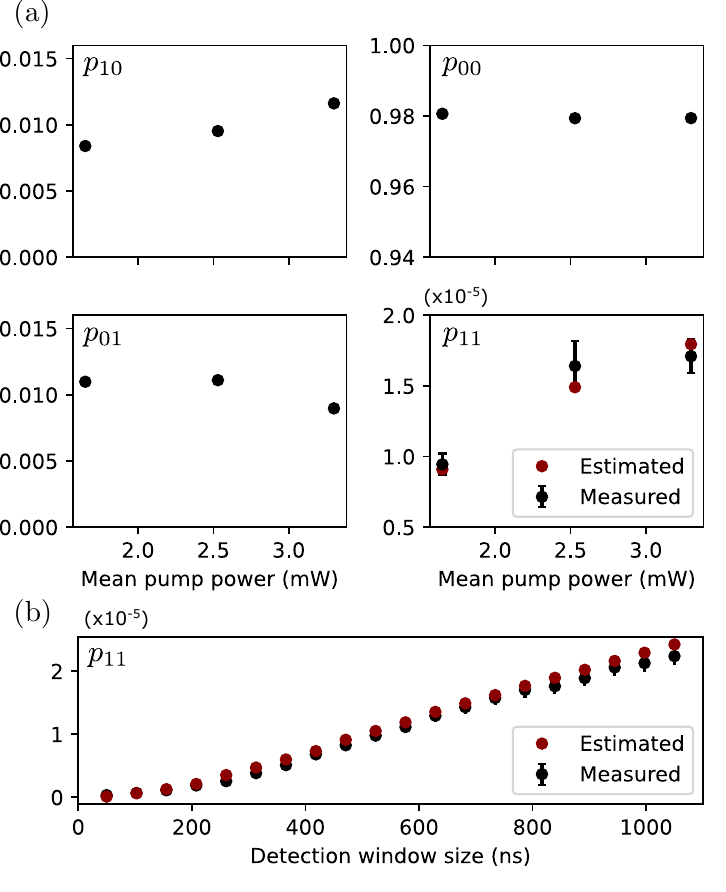}
    \caption{(a) Experimentally measured diagonal elements of $\rho$ without storage (black circles). Here, the QMs are prepared with a wide transparency window thus increasing the three-fold coincidence rate and allowing for the verification of the model for $p_{11}$ given in~\cref{eq:p11_model}. Red markers correspond to the estimated values of $p_{11}$. Error bars are calculated assuming poissonian statistics and detection window size is set to \SI{800}{\nano\second}. The x-axis corresponds to the average pump power measured at the two cSPDC sources. (b) Estimated and measured $p_{11}$ as a function of detection window size.}
    \label{fig:p11_model}
\end{figure}
Owing to the very low three-fold coincidence rate of our experiment, the $p_{11}$ element of the density matrix had to be estimated from the measurement of $p_{10}$ and $p_{01}$. Indeed, with a reported value of \num{5.9(1)d-8} in Table~\ref{tb:diags} and a total heralding rate of \SI{510}{\cps} in conditional storage, approximately \num{0.1} three-fold coincidence events are measured every hour, making the required acquisition time needed to have reasonably low uncertainty on $p_{11}$ very large. 

To circumvent this issue, we devise a model in which we divide the two-fold event probabilities into the sum of the true coincidences and the accidental noise background $p_{ij} = p_{\text{coinc},ij} + p_{\text{acc},ij}$. We then estimate the three-fold probability by assuming that triple coincidences can arise from recording simultaneously, upon idler detection, either a true photon detection event and an accidental or two accidentals. This leads to the following expression for $p_{11}$:
\begin{equation}\label{eq:p11_model}
    p_{11} = p_{\text{coinc},10} p_{\text{acc},01} + p_{\text{acc},10} p_{\text{coinc},01} + p_{\text{acc},10} p_{\text{acc},01}.
\end{equation} 
We point out that the model is formally equivalent to the one used in ref.~\cite{Lago-Rivera2021} although formulated in terms of true and accidental coincidences rather than isolated node cross-correlations.
The values of $p_{\text{coinc},10},p_{\text{coinc},01}$ are obtained by subtracting the measured accidental counts to the experimental values of $p_{10},p_{01}$. Accidentals are obtained from the corresponding time-correlation histograms by averaging the recorded counts over two broad \SI{2}{\micro\second} located before and after the \SI{280}{\nano\second}-wide coincidence window, as shown in \cref{fig:g2s}~(b-c). The use of two noise windows of large duration guarantees a faithful estimation of the noise background inside the echo temporal mode. We emphasize that our model is general and does not require any assumption on the origin of accidental counts, which can be caused by CP fluorescence, scattering of stray light, spurious free induction decay or double pair emission.

To corroborate the validity of our model, we perform a series of additional measurements where the diagonal elements of $\rho$ are measured in the absence of storage, with a transparency window prepared in the two QMs. In this setting the increased count rate allows for direct measurement of $p_{11}$ and comparison with the value derived from~\cref{eq:p11_model}. We perform this measurement for different values of the source pump power and with the detection window size set to \SI{800}{\nano\second}. The results are displayed in~\cref{fig:p11_model}~(a), showing excellent agreement between the directly measured and estimated values of $p_{11}$.   

To further validate the model, we compare the estimated and measured values of $p_{11}$ as a function of the detection window size. The results are plotted in~\cref{fig:p11_model}~(b), showing good agreement between the two. For window sizes above \SI{200}{\nano\second}, we observe that the estimated value is slightly above the measured one, meaning our model is giving a pessimistic estimate of the concurrence $\mathcal{C}$. We also note that the quadratic scaling of $p_{11}$ for narrow detection windows is well captured by the model. The explanation for this scaling is discussed in more details in the following section.

\subsection{Scaling of the concurrence with detection window size}

\begin{figure}
    \centering
    \includegraphics[width=\linewidth]{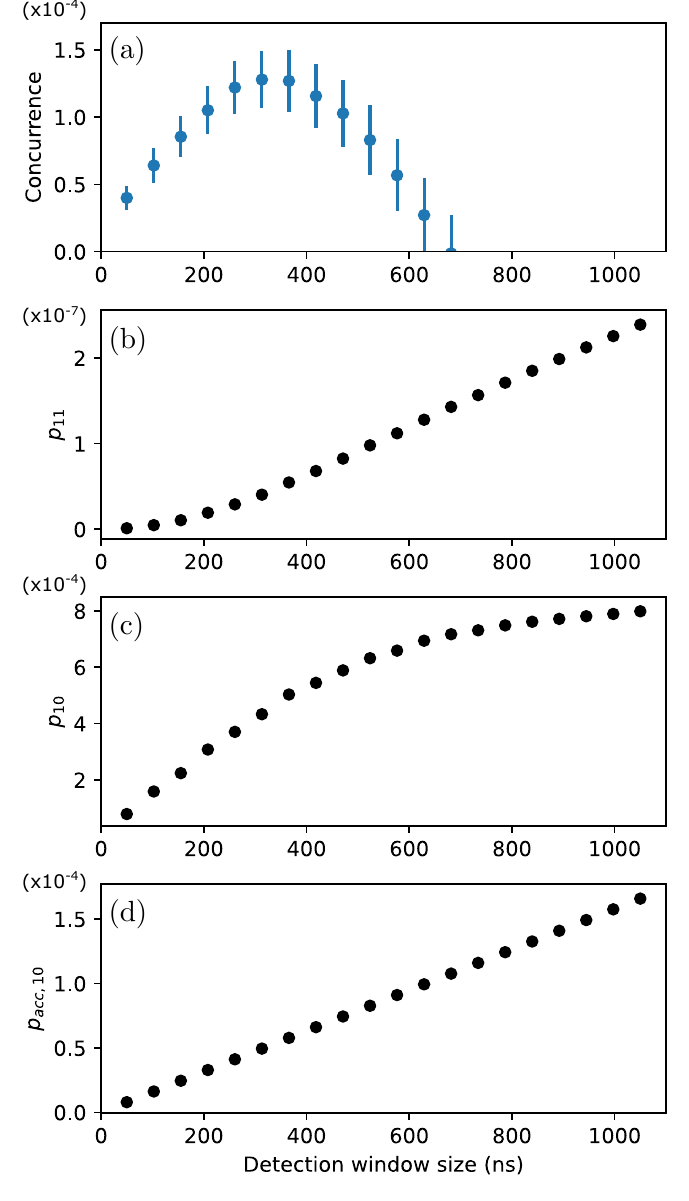}
    \caption{(a) Concurrence, (b) $p_{11}, $(c) $p_{10}$ and (d) $p_\text{acc,10}$} as a function of detection window size for the dataset featured in Section~\ref{sec:ff} (measurement with phase feed-forward).
    \label{fig:concurrence_detwin}
\end{figure}

For all the data presented in the main text of this article, the coincidence window size $w$ has been set to \SI{280}{\nano\second}. In~\cref{fig:concurrence_detwin} we show the concurrence $\mathcal{C}$ as well as the probabilities $p_{11}, p_{10}$ and $p_\text{acc,10}$ for the dataset featured in Section~\ref{sec:ff} as a function of $w$. As can be seen in~\cref{fig:concurrence_detwin}~(a), the concurrence exhibits a maximum at around \SI{300}{\nano\second}, making our choice of window size optimal for maximizing the degree of measured entanglement. 

Nevertheless, the decrease of the concurrence for small window size can appear rather counter-intuitive. Indeed, when decreasing $w$ the value of $g^{(2)}_{si}$ and therefore of $V$ increases, according to \cref{eq:vis_est}. On the other hand, one would expect $p_{11}$ to keep decreasing with $w$ at a greater rate than that of $p_{10},p_{01}$ resulting in an overall increase of $\mathcal{C} = \max(0,V(p_{10}+p_{01})-2\sqrt{p_{00}p_{11}})$. However, as displayed in~\cref{fig:concurrence_detwin}~(b), the value of $p_{11}$ shows an inflection when approaching $w=0$ which in turns results in a slower decrease of the negative term in $\mathcal{C}$, explaining the loss of entanglement for small detection window. This can be explained qualitatively from the model of \cref{eq:p11_model} assuming $p_{10}=p_{01}$ and $p_\text{acc,10} = p_\text{acc,01}$. For small $w$, we can consider $p_\text{acc,10}$ to be small and therefore $p_{11} \approx 2p_{10}p_\text{acc,10}$. From~\cref{fig:concurrence_detwin}~(c-d) we see that, in the vicinity of $w=0$ $p_{10}$ and $p_\text{acc,10}$ are linear in $w$ which means that $p_{11}$ scales quadratically with the window size. This quadratic scaling explains the measured inflection in $p_{11}$ and consequently the decrease in $\mathcal{C}$.

We emphasize that this scaling of $p_{11}$ with window size is not an artifact of the model used to estimate this quantity, since it can also be observed in a direct measurement without storage in the quantum memory, as shown in \cref{fig:p11_model}~(b).

\subsection{Scaling with pump power}
\label{sec:swg2model}

As stated in the main text, the amount of available pump power was limited by the nominal output of the diode laser to under \SI{4}{\milli\watt} per source. In this appendix, we show that this is merely a technical limitation and that we could use larger pump powers to increase the heralding rates while still maintaining positive concurrence. 

To begin with, we give a lower bound on the value of $g^{(2)}_{si}$ to keep the concurrence $\mathcal{C}$ above zero. From the definition, we have $\mathcal{C} = \max(0,2d-2\sqrt{p_{00}p_{11}})$. Following ref.~\cite{Lago-Rivera2021} and assuming that $p_{10}=p_{01}$ and $p_\text{acc,10} = p_\text{acc,01}$, we estimate $p_{11}$ from the relation \eqref{eq:p11_model} as 
\begin{equation}\label{eq:p11g2}
p_{11} = 4\left(\frac{p_{10}}{g^{(2)}_{si}}\right)^2(g^{(2)}_{si}-1),
\end{equation}
where we consider that the value of $g^{(2)}_{si}$ is the average over the two nodes. Then, we rewrite eq.~ \eqref{eq:vis_est} in the form
\begin{equation}\label{eq:visg2}
    V=V_\text{lim}\frac{g^{(2)}_{si}-1}{g^{(2)}_{si}+1}.
\end{equation}
By combining eqs.~\eqref{eq:p11g2} and~\eqref{eq:visg2} with the definition of the concurrence we show that for keeping $\mathcal{C}>0$ the following inequality must hold:
\begin{equation}\label{eq:ineqg2}
   \frac{g^{(2)}_{si}\sqrt{g^{(2)}_{si}-1}}{2\left(g^{(2)}_{si}+1\right)}> \frac{\sqrt{p_{00}}}{V_\text{lim}}.
\end{equation}
From the numerical values reported in Appendix~\ref{sec:vis_est}, we obtain $V_\text{lim}=\SI{67(7)}{\percent}$ and consequently estimate that a minimum value of $g^{(2)}_{si}=16$ is required to ensure positive concurrence by two standard deviations.

We then calculate the expected value of the cross-correlation $g^{(2)}_{si}$ of a single memory measured after spin-wave storage using the following expression~\cite{Maring2018}:
\begin{equation}\label{eq:g2simodel}
    g^{(2)}_{si} = g^{(2)}_\text{AFC} \dfrac{\eta_H/\mu_1 + 1}{\eta_H/\mu_1+ g^{(2)}_\text{AFC}},
\end{equation}
with $g^{(2)}_\text{AFC}$ the cross-correlation measured with AFC only in the absence of CPs, $\eta_H$ the heralding efficiency of the cSPDC source and $\mu_1$ the number of input photons required for attaining a signal-to-noise ratio of 1. Our system features typical values of $\eta_H=\SI{20}{\percent}$ and $\mu_1=\SI{1}{\percent}$ while $g^{(2)}_\text{AFC}$ has been measured to be \num{92 +- 19} at \SI{3.55}{\milli\watt} pump power for Node A and \num{157+-32} at \SI{4.2}{\milli\watt} pump power for Node B at a storage time of $\tau_\text{AFC}=\SI{10}{\micro\second}$. This yields estimated values for $g^{(2)}_{si}$ of \num{17(5)} and \num{19(5)} for Node A and B respectively, in good agreement with the measured values reported in Fig.~\ref{fig:g2s}~(b-c). 

To show the evolution with pump power, we point out that the only quantity in eq.~\eqref{eq:g2simodel} that depends on it is $g^{(2)}_\text{AFC}$ which exhibits a inverse-law scaling: $g^{(2)}_\text{AFC}=1+1/aP$ with $P$ the pump power and $a$ a proportionality constant (see Supplementary Material of ref.~\cite{Lago-Rivera2021}). When $g^{(2)}_\text{AFC}$ is large, its value is thus inversely proportional to $P$. 
We calculate that the limit value of $g^{(2)}_{si}=\num{16}$ for a positive concurrence derived from equation \eqref{eq:ineqg2} is reached when multiplying the pump power per source by a factor \num{3.9}. In other words, we have shown that by using a more powerful pump laser our system would allow for a nearly fourfold increase of the heralding rate while still maintaining entanglement between the two QMs.

\end{document}